\documentclass[journal=nalefd,manuscript=article,layout=twocolumn]{achemso}

\usepackage{chemformula} %
\usepackage[T1]{fontenc} %
\usepackage{xcolor}
\usepackage{amssymb}
\definecolor{cubic_diamond}{HTML}{13A0FE}
\definecolor{cubic_diamond_1st}{HTML}{00FEF5}
\definecolor{cubic_diamond_2nd}{HTML}{7EFEB5}
\definecolor{hex_diamond}{HTML}{FE8900}
\definecolor{hex_diamond_1st}{HTML}{FEDC00}
\definecolor{hex_diamond_2nd}{HTML}{CCE551}

\author{Hongjin Du}
\affiliation[Cornell University]
{Department of Materials Science and Engineering, Cornell University, Ithaca, New York, 14853, United States}
\author{Ellery J.\ Hendrix}
\affiliation[Cornell University]
{Department of Materials Science and Engineering, Cornell University, Ithaca, New York, 14853, United States}
\altaffiliation{Current address: Department of Materials Science and Engineering, University of Michigan, Ann Arbor, Michigan, 48109, United States}
\author{Richard D.\ Robinson}
\affiliation[Cornell University]
{Department of Materials Science and Engineering, Cornell University, Ithaca, New York, 14853, United States}
\author{Julia Dshemuchadse}
\email{jd732@cornell.edu}
\affiliation[Cornell University]
{Department of Materials Science and Engineering, Cornell University, Ithaca, New York, 14853, United States}

\title[MSC]{Understanding the Structural Origin of Chirality in Magic-Size Semiconductor Nanoclusters through Self-Assembly Simulations}

\begin{document}


\begin{abstract}
Semiconductor magic-size clusters (MSCs) are atomically precise nanoparticles exhibiting unique size-dependent properties, but their ultrasmall dimensions hinder structural characterization, limiting our understanding of their formation and stability. A few MSC structures have been fully resolved, revealing either bulk-like zincblende-type structures or a range of non-bulk-like motifs. 
Here we use a computational model to investigate the relationship between cluster size and atomic structure in zincblende-forming %
II–VI and III–V semiconductors. 
Firstly, we find that all non-bulk-like MSCs in these systems exhibit the same distorted icosahedral motif that is intrinsically chiral. 
Secondly, we reproduce these MSC geometries in small-cluster self-assembly simulations and discover
that their chirality emerges from the geometric frustration and symmetry breaking in arranging tetrahedral bonding environments into an icosahedral topology. Overall, this work reproduces experimentally reported motifs without system-specific parameterization, establishes the structural origin of chirality in MSCs, and provides design principles for predicting new cluster geometries.
\end{abstract}

\section{Introduction}

Colloidal semiconductor magic-size clusters (MSCs) are atomically precise nanoclusters composed of tens to a few hundred atoms, occupying a size regime between molecules and nanocrystals \cite{soloviev_size-dependent_2001, gary_single-crystal_2016, williamson_chemically_2019, bootharaju_structure_2022, chen_semiconductor_2025}. These clusters represent kinetically stabilized intermediates with enhanced (meta)stability compared to neighboring sizes. Their growth is quantized and non-classical, frequently appearing as discrete intermediates in nanocrystal synthesis \cite{kasuya_ultra-stable_2004, kudera_sequential_2007, beecher_atomic_2014, lee_nonclassical_2016}. 
Due to their size uniformity, MSCs offer a promising route to address the challenge of size polydispersity in colloidal nanocrystal synthesis \cite{talapin_dynamic_2002, pu_colloidal_2018}. MSCs have also emerged as versatile building blocks for constructing functional suprastructures. 
For instance, the periodic assembly of Mn$^{2+}$-doped (CdSe)$_{13}$ and (ZnSe)$_{13}$ MSCs into 3D and 2D superstructures have shown improved stability, enhanced solid-state photoluminescence quantum yields, and increased catalytic activity \cite{baek_highly_2021}. Suprastructures of (CdSe)$_{13}$ have also demonstrated promising photocatalytic properties for hydrogen production \cite{lee_photocatalytic_2025}.
In another example, CdS MSCs were shown to assemble through meniscus-guided evaporation into hierarchically ordered chiral films spanning six orders of magnitude in length scale \cite{han_multiscale_2022}. This approach was later generalized to CdSe and CdTe MSCs, producing materials with exceptionally strong circular dichroism. \cite{ugras_transforming_2025}.

Among II–VI and III–V semiconductor MSCs, which have been most extensively studied, resolved crystal structures tend to fall into two categories. 
The first type adopts a structure that exhibits a zincblende-type topology and tetrahedral cluster shapes, resulting in clusters that resemble fragments of their bulk structure \cite{dance_syntheses_1984, lee_s4cd17sph282-_1988, herron_crystal_1993, vossmeyer_double-diamond_1995, vossmeyer_double-layer_1995, jin_synthesis_1996, behrens_synthesis_1996, eichhofer_thermal_2005, beecher_atomic_2014}. 
The second type of clusters, while also exhibiting local tetrahedral motifs, does not exhibit bulk-like topologies and typically adopts quasi-spherical cluster shapes. In the literature, these have been variously described as “wurtzite-like” or simply as lacking clear bulk analogues \cite{khadka_zinc_2012, gary_single-crystal_2016, bootharaju_structure_2022, ma_precision_2023, sandeno_ligand_2024, sandeno_synthesis_2024, xu_chiral_2025}. 

Despite their structural diversity, we find that the non-bulk-like MSCs in these semiconductor systems share a common underlying motif: a core of a distorted icosahedral framework with $X_{14}Y_{13}$ composition. This motif has been described as ``multiple polytwistane units helically intersected'' as in the In$_{13}$P$_{14}$ core of In$_{37}$P$_{20}$ \cite{ritchhart_templated_2018} and as an ``$A_{14}B_{13}$ skeleton with 13 anions assembled into an icosahedron'' \cite{ma_programmable_2025}. 
This recurring motif allows us to unify non-bulk-like MSCs within a single structural category, which we examine in detail below.

A notable yet underappreciated feature of these non-bulk-like MSCs is their chirality. While recent studies have explicitly reported chiral MSCs---such as Cd$_{26}$Se$_{17}$ \cite{ma_precision_2023} and Cd$_{28}$S$_{17}$ \cite{xu_chiral_2025} synthesized \textit{via} cation-exchange reactions---this work will show that other MSCs, including Zn$_{14}$S$_{13}$ \cite{khadka_zinc_2012}, Cd$_{14}$Se$_{13}$ \cite{bootharaju_structure_2022}, In$_{37}$P$_{20}$ \cite{gary_single-crystal_2016}, In$_{26}$P$_{13}$ \cite{sandeno_ligand_2024}, and In$_{26}$As$_{18}$ \cite{sandeno_synthesis_2024}, are also inherently chiral, even though chirality was not discussed in the original reports. 
This suggests that chirality is not an isolated feature, but rather an intrinsic characteristic of this class of MSCs.

Despite the growing number of experimentally resolved MSC structures, the origins of chirality 
remain poorly understood. 
Prior theoretical studies have primarily aimed to predict the lowest-energy configurations and size-dependent properties of MSCs using density functional theory (DFT), sampling the energy landscape \textit{via} Monte Carlo-based global optimization \cite{farrow_structure_2014}, a combination of structure enumeration with Monte Carlo search and local optimization \cite{nguyen_understanding_2010,nguyen_computational_2013}, \textit{ab-initio} random structure searching (AIRSS) \cite{tan_structures_2019,zhang_structure_2025}, simulated annealing \cite{lopez-morales_exploring_2020}, and unbiased fuzzy global optimization \cite{lei_structural_2023}. 
Subsequent calculations of optical excitations, when compared with experimental spectra, validated candidate structures with limited success. \cite{botti_identification_2007,nguyen_understanding_2010,del_ben_density_2011,nguyen_computational_2013}.  
In practice, however, these strategies tend to produce extensive libraries of low-energy minima, often failing to consistently reproduce experimentally resolved clusters, which might also be metastable or stabilized by competing interactions in the system. 
Additionally, predictions are often system-specific and sometimes mutually inconsistent---for example in the stoichiometric (CdSe)$_n$ system, where the proposed structures remain experimentally unverified and no consensus has emerged \cite{zhang_structure_2025}. 
The predicted structures and spectra are also highly sensitive to the choice of the DFT functional, with different functionals favoring distinct structural motifs and yielding varying accuracy in excitation energies \cite{nguyen_computational_2013,sigalas_size_2014}. 
As a result, neither the structural origin of chirality in MSCs nor a unifying link to the distorted icosahedral motif commonly reported for this class has yet been established.

Here we present a computational model that captures the geometric features of non-bulk-like MSC self-assembly. This model enables us to reproduce and investigate the emergence of specific structural arrangements observed in experimentally resolved MSC structures. Using a chemically agnostic pair potential, we simulate the self-assembly of ligand-free nanoclusters across a wide range of sizes and stoichiometries. Our method captures the transition from achiral zincblende-type bulk structures at larger sizes to chiral, non-bulk-like motifs at smaller sizes---mirroring trends observed in experimentally studied MSCs.

Through a comprehensive structural analysis we find that the emergence of the chiral distorted icosahedral motifs is due to an underlying geometric frustration and symmetry reduction, which offers a unifying framework for interpreting the symmetry-breaking features observed across these diverse II–VI and III–V systems.

\section{Method}

Our approach to modeling semiconductor MSCs is guided by the crystal structures observed in the corresponding bulk materials: group II–VI and III–V semiconductors form zincblende or wurtzite phases, both tetrahedrally coordinated, and differing only in their stacking patterns \cite{yeh_1992} (see Fig.~S2 in the Supporting Information).
We employ oscillatory pair potentials (OPPs) to model atomic interactions in these systems \cite{mihalkovic_empirical_2012}. These potentials explicitly incorporate Friedel oscillations, modulations of electron density that arise from charge screening by conduction electrons at the Fermi surface \cite{friedel_electronic_1954,friedel_metallic_1958}. This behavior is physically equivalent to the Hume-Rothery mechanism \cite{pettifor_bonding_1995}, which links the electronic structure of metals to the stability of specific crystal arrangements. 
Over recent years, OPPs and other isotropic pair potentials have been shown to mimic the structure formation behavior of a wide range of materials.\cite{Dshemuchadse2021,Pan2023,Du2025,Pan2025}

\begin{figure*}
  \includegraphics[width=\textwidth]{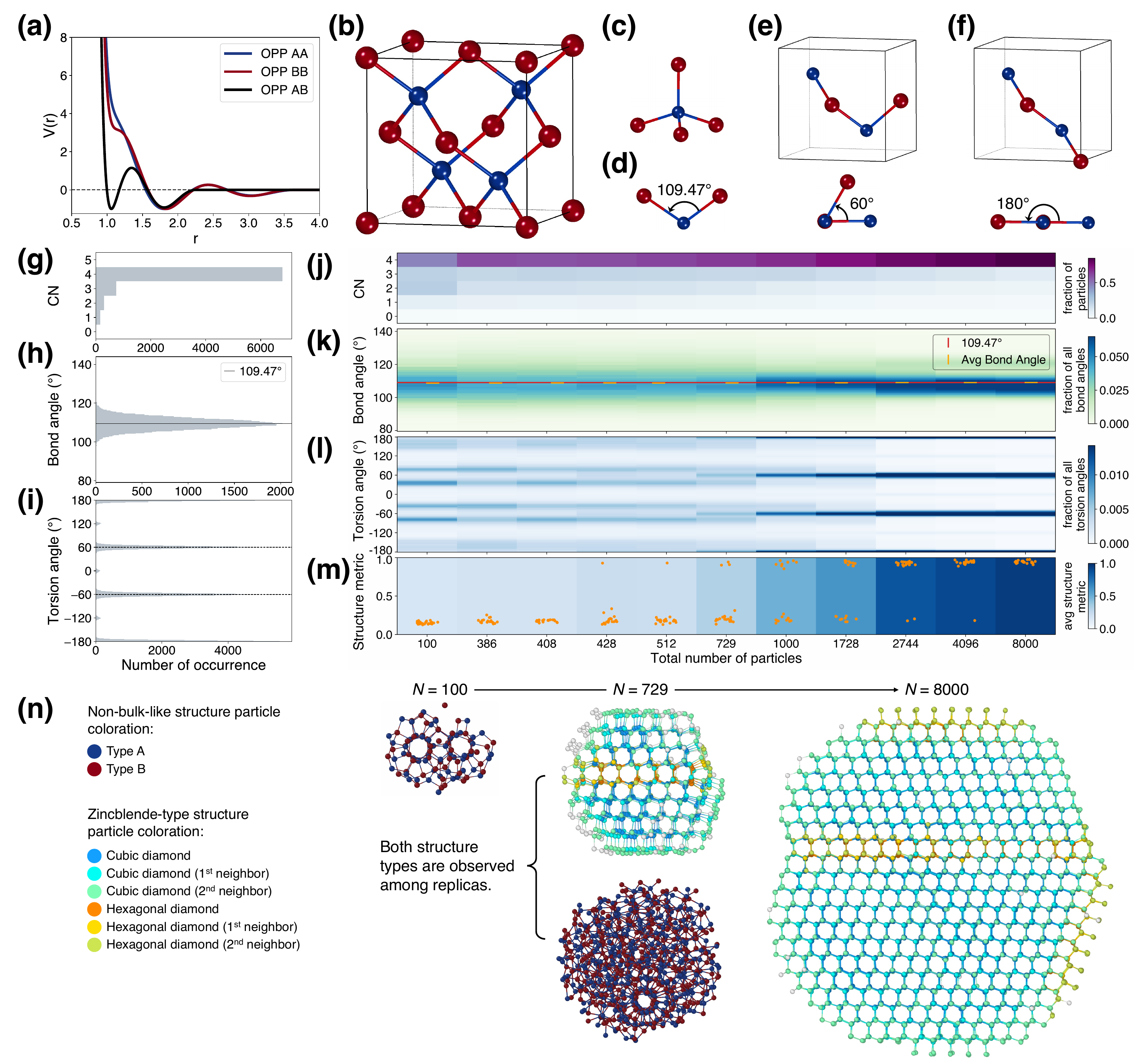}
  \caption{\small Computational model and structural analysis of self-assembled, stoichiometric bulk and cluster configurations. 
  \textbf{(a)} Isotropic pair potentials for particle interactions in the binary system: $A$–$A$ (blue), $B$–$B$ (red), and $A$–$B$ (black). 
  \textbf{(b)} Unit cell of the zincblende-type structure, with particle types $A$ and $B$ shown in blue and red, respectively. 
  The zincblende-type structure has \textbf{(c)} coordination number (CN) = 4, \textbf{(d)} bond angles 109.47°, and torsion angles \textbf{(e)} ±60° and \textbf{(f)} 180°. 
  Bulk simulations (with $N = 8000$ particles) confirm the formation of a zincblende-type structure: \textbf{(g)} the CN distribution peaks at four, \textbf{(h)} the bond angle distribution is centered around 109.47°, and \textbf{(i)} torsion angles are concentrated at ±60° and 180°.
  For clusters with $N=100$ to 8000 particles, \textbf{(j)} the fraction of CN = 4 sites and \textbf{(k)} the average bond angle converge toward the ideal zincblende values as cluster size increases. \textbf{(l)} Torsion angle distributions evolve from non-bulk-like to bulk-like patterns (with peaks at ±60° and 180°), which is reflected in \textbf{(m)} the structure metric for each replica (orange dots) shifting from values below 0.35 to above 0.85. 
  \textbf{(n)} Representative structures at different cluster sizes demonstrate that: 
  small systems ($N = 100$) exhibit only non-bulk-like structures, 
  intermediate-size systems ($N = 729$) can form either zincblende-type structure or non-bulk-like structures, and 
  large systems ($N = 8000$) adopt a bulk-like zincblende-type structure. 
  (Particles in zincblende-type structures are colored according to the ``Identify diamond structure'' order parameter with \textit{OVITO}.\cite{maras_global_2016})
  }
  \label{fig:opp}
\end{figure*}

From a high-dimensional search over binary mixtures using a machine learning-based optimization framework \cite{ML_potential}, we select a set of OPP parameters (Fig.~\ref{fig:opp}a) that model the formation of the zincblende structure in bulk simulations. \cite{ML_potential}
The model comprises two particle species, $A$ and $B$, with three distinct pair interactions: the unlike-particle ($A$--$B$) interaction potential exhibits an attractive minimum at interparticle distances $r \approx 1.1$, whereas the same-type-particle ($A$--$A$ and $B$--$B$) potentials are only attractive at the next-nearest-neighbor distance ($r \approx 1.8$). These interaction patterns promote effectively salt-like behavior and the formation of the zincblende structure type.

We model the assembly of these systems with molecular dynamics simulations, cooling each system from a low-density, high-temperature gas phase down to low temperatures using a linear temperature ramp in the $NVT$ ensemble (with a Nos\'e--Hoover thermostat, $\delta t = 0.005$). 
We simulate kinetically accessible cluster configurations with varying numbers of building blocks and varying stoichiometries, 
and we conduct replica simulations at all state points to gather statistics. 
It is important to note that the finite-size assemblies in our simulations do not arise from self-limiting growth; instead, their size is predetermined by the number of particles specified at initialization.

To quantify the local coordination geometry and three-dimensional topology of the assembled clusters, we calculate the coordination number and bond angles for each particle from its nearest neighbors. In addition, we compute torsion angles for bonded quadruplets, defined by four sequentially bonded and distinct particles $i$–$j$–$k$–$l$. Each torsion angle corresponds to the angle between the planes defined by particles $(i, j, k)$ and $(j, k, l)$ (see Fig.~S1 in the Supporting Information). 

All simulations are performed using \textit{HOOMD-blue} \cite{HOOMD} and visualized with \textit{CrystalMaker}\textsuperscript{\textregistered} \cite{palmer2024} and \textit{OVITO} \cite{ovito}; 
structural analysis is conducted with the \textit{freud} software package \cite{Freud} and custom-written code. 
Further details on simulation parameters and analysis protocols are provided in the Supporting Information.

\section{Results and discussion}

Consistent with the structural characteristics of the zincblende structure type (Figure~\ref{fig:opp}b--f), the simulated bulk configurations exhibit coordination numbers of predominantly CN $= 4$ (Figure~\ref{fig:opp}g), bond angles centered around the tetrahedral angle of 109.47° (Figure~\ref{fig:opp}h), and torsion angles peaking at ±60° and 180° (Figure~\ref{fig:opp}i).
As we decrease the system size to simulate nanocluster formation, the self-assembled structures progressively deviate from ideal zincblende values corresponding to bulk-like characteristics, ultimately exhibiting non-bulk-like motifs with differing coordination environments and bond geometries (Figure~\ref{fig:opp}j--l). 
Specifically, the fraction of tetrahedrally coordinated particles (CN $= 4$) decreases (Figure~\ref{fig:opp}j), the average bond angle falls below 109.5° (Figure~\ref{fig:opp}k), and the torsion angle distribution shifts from the bulk-like peaks at ±60° and 180° to six distinct peaks at different angular positions (Figure~\ref{fig:opp}l).

We introduced a simple structure metric to distinguish the zincblende structure type from non-bulk-like structures, defined as the fraction of torsion angles falling near the zincblende reference peaks at ±60° and 180° (see Supporting Information for details). 
Small clusters ($N \leq 408$) exhibit low values for the structure metric ($<0.35$) indicative of non-bulk-like structures, while the proportion of replicas with high values ($>0.85$), reflective of the zincblende structure, increases systematically with system size (Figure~\ref{fig:opp}m, orange dots).
This result indicates that at large sizes ($N = 8000$), all replicas are zincblende-like, while at small sizes ($N \leq 408$) all replicas are non-bulk-like.
In the intermediate regime ($428 \leq N \leq 4096$), zincblende and non-bulk-like motifs coexist, with the fraction of zincblende-type replicas decreasing as system size decreases (Figure~\ref{fig:opp}j--n and Fig.~S3 in the Supporting Information).

\begin{figure}
  \includegraphics[width=0.5\textwidth]{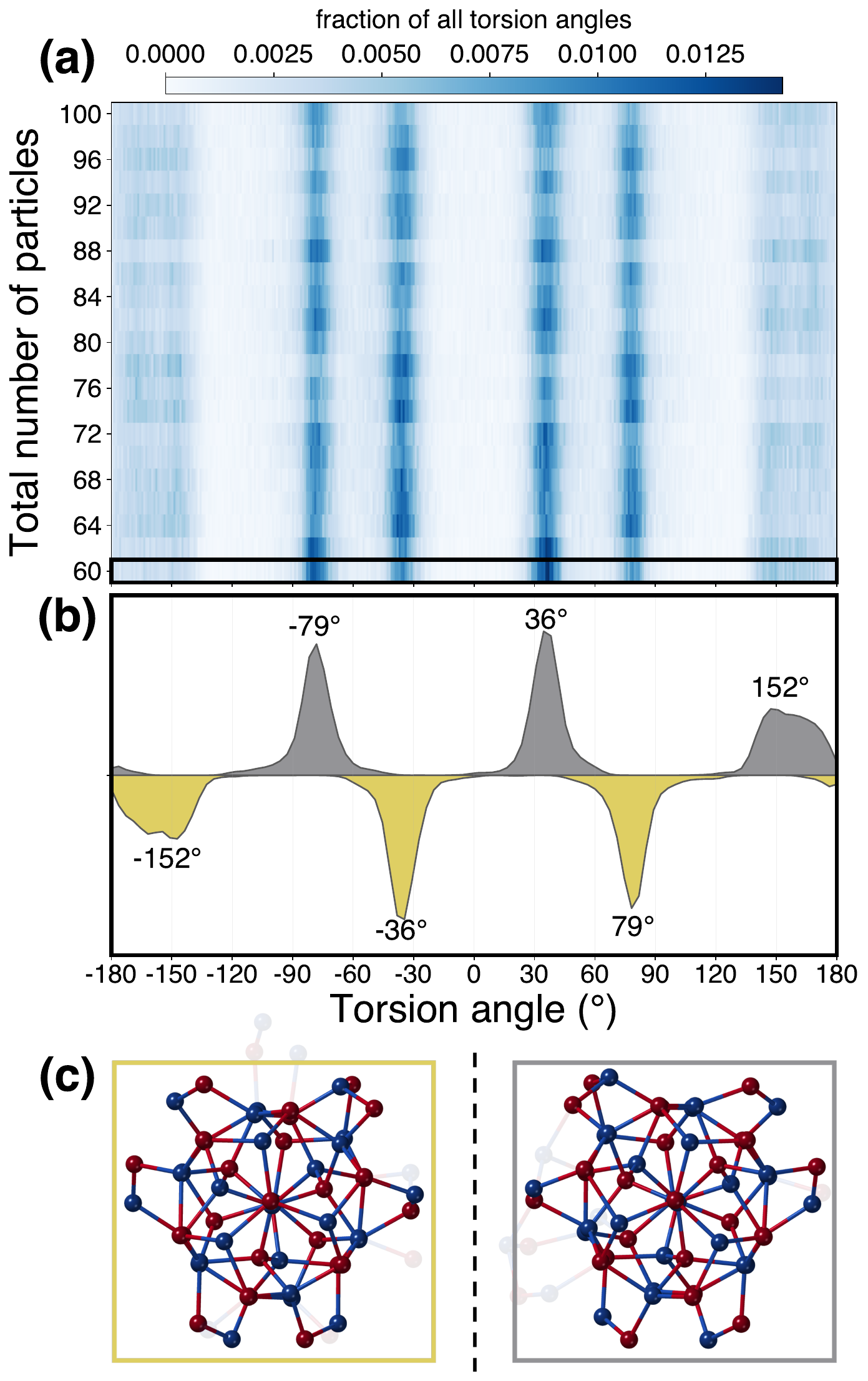}
  \caption{Torsion angle distributions and representative structures for clusters at 1:1 stoichiometry. 
  \textbf{(a)} Torsion angle distribution for cluster sizes $N = 60$--100, averaged over 30 replicas each. 
  \textbf{(b)} Torsion angle distribution of 30 replicas at $N = 60$, grouped by chirality and normalized by number of replicas (18 replicas in gray, 12 in yellow). Peaks at $-79$\textdegree / $+36$\textdegree / $+152$\textdegree\ and $-152$\textdegree / $-36$\textdegree / $+79$\textdegree\ indicate enantiomers of a chiral species. 
  \textbf{(c)} Representative structures at $N = 60$ showing opposite handedness (opaque particles denote configurations of opposite chirality).}
  \label{fig:TA}
\end{figure}

Building on these size-dependent trends, we next focus on the small systems---clusters with ideal zincblende stoichiometry (with a 1:1 ratio of $A$- to $B$-type particles, \textit{i.e.}, with $A_{N/2}B_{N/2}$ stoichiometry) and $N$ = 60–100 particles---to examine how their local geometries depart from bulk-like order and give rise to distinct torsional angle patterns.
Across this cluster-size range, the torsion-angle distributions are essentially identical (Figure~\ref{fig:TA}a), yet they notably deviate from the bulk zincblende structure, showing distinct peaks at around ±36°, ±78°, and ±158° (with minor variations depending on $N$, see also Fig.~S9 in the Supporting Information), instead of the characteristic ±60° and 180°. This shift signals a significant deviation from the zincblende structure type, suggesting the emergence of distinct non-bulk-like motifs.
While the averaged distributions across replicas contain all six torsion-angle peaks, individual replicas consistently present only one of two complementary sets---for example, +36°/–79°/+152° or –36°/+79°/–152° at $N$ = 60 (Figure~\ref{fig:TA}b)---indicating the presence of two enantiomorphic forms in the nanoclusters \cite{Klyne1960}.

Visualizing individual clusters confirms that replicas with reverse torsion-angle patterns adopt structures of opposite handedness (Figure~\ref{fig:TA}c). 
Despite this intrinsic chirality at the single-cluster level, the averaged torsion-angle distributions across all replicas are symmetric, meaning that the simulations produce a racemic mixture of left- and right-handed clusters with no global preference for either handedness in this size range. 
The torsion-angle distributions document that clusters are structurally stable and retain their handedness upon moderate heating, with no inversion of chirality observed below the crystallization temperature (see Figs.~S15 and S16 in the Supporting Information).

\begin{figure*}
  \includegraphics[width=\textwidth]{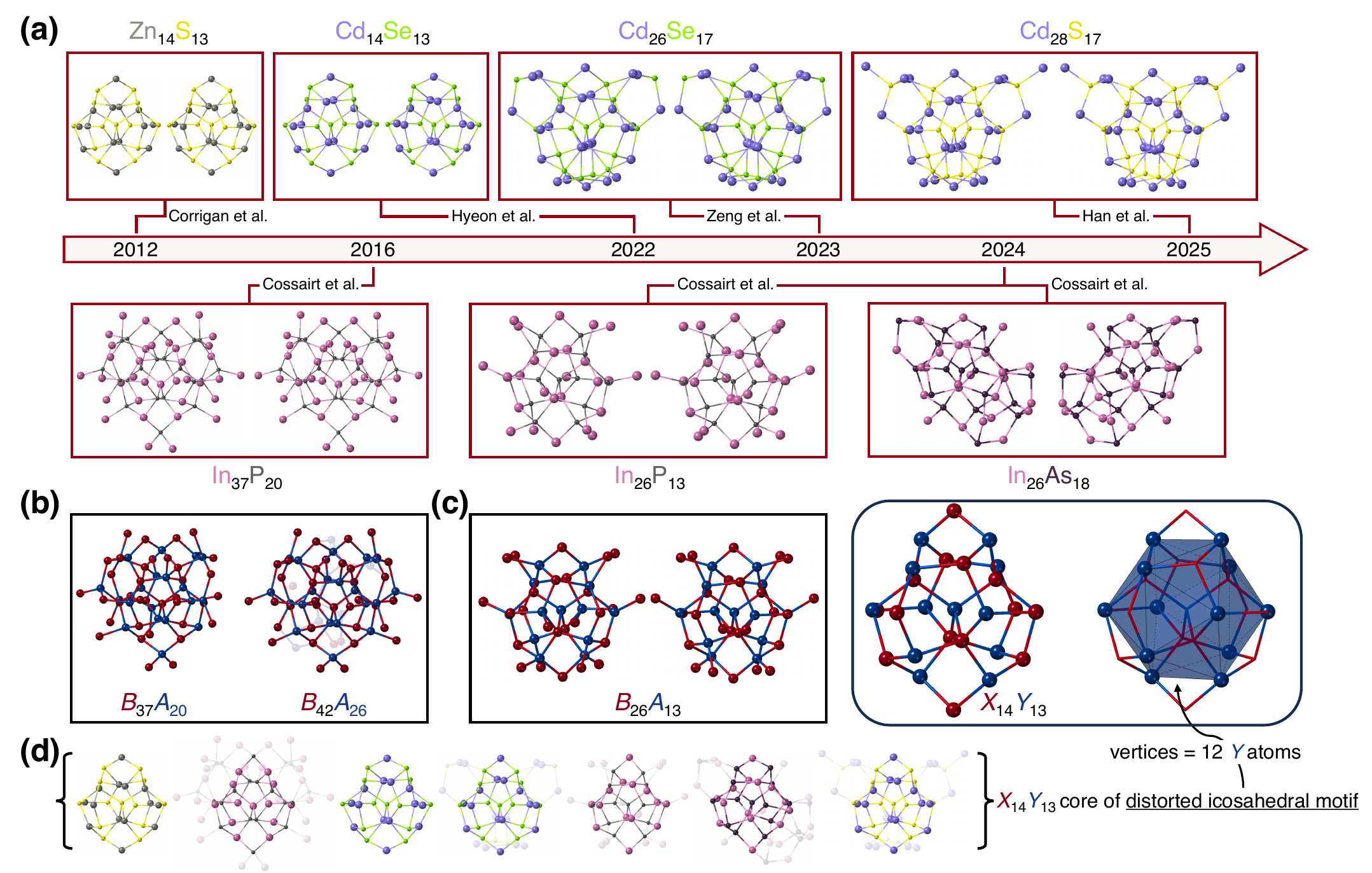}
  \caption{Shared distorted icosahedral motif across experimentally observed chiral MSCs and simulated cluster analogs, and mechanism of chirality emergence. 
  \textbf{(a)} Timeline of the discovery of chiral semiconductor MSCs, showing the core structures of representative compositions. For each cluster type, both chiral enantiomers are displayed side by side. Shown, chronologically from the earliest to most recent reports, are: Zn$_{14}$S$_{13}$ \cite{khadka_zinc_2012}, In$_{37}$P$_{20}$ \cite{gary_single-crystal_2016}, Cd$_{14}$Se$_{13}$ \cite{bootharaju_structure_2022}, Cd$_{26}$Se$_{17}$ \cite{ma_precision_2023}, In$_{26}$P$_{13}$ \cite{sandeno_ligand_2024}, In$_{26}$As$_{18}$ \cite{sandeno_synthesis_2024}, and Cd$_{28}$S$_{17}$ \cite{xu_chiral_2025}.
  \textbf{(b--c)} Comparison of simulated nanoclusters with experimentally observed chiral MSCs: 
  \textbf{(b)} simulated $B_{37}A_{20}$ and the $B_{37}A_{20}$ core in a $B_{42}A_{26}$ cluster (with shell particles shown translucently), resembling the In$_{37}$P$_{20}$ shown in (a), and  
  \textbf{(c)} simulated $B_{26}A_{13}$ cluster, sharing the same topology as the In$_{26}$P$_{13}$ shown in (a).
  \textbf{(d)} Shared structural motif across both experimental and simulated chiral clusters: a distorted icosahedral motif composed of 14 $X$ particles and 13 $Y$ particles. Atoms are colored the same as in (a) and are shown in the same chronological order from the earliest to the most recent reports.}
  \label{fig:msc_structures}
\end{figure*}

We compared the chiral, non-bulk-like MSCs observed in our simulations with 
a range of experimentally observed clusters, reported in literature, whose structures had been resolved by single-crystal X-ray diffraction and exhibited non-bulk-like structures (Figure~\ref{fig:msc_structures}a). 
A detailed structural analysis of these experimental non-bulk-like MSCs is provided in the Supporting Information.

In our simulations, the $B_{37}A_{20}$ core of the slightly larger $B_{42}A_{26}$ cluster is topologically identical to the In$_{37}$P$_{20}$ cluster from literature \cite{gary_single-crystal_2016}, with a mean absolute error (MAE) of 5.31° across corresponding bond angles in the two structures (see Figs.~S11 and S13 in the Supporting Information). Here, the MAE quantifies structural similarity as the average of the absolute differences between corresponding bond angles in the two clusters.
The $B_{37}A_{20}$ cluster also resembles the In$_{37}$P$_{20}$ structure \cite{gary_single-crystal_2016}, though with subtle differences in the arrangement of a few surface particles (Figure~\ref{fig:msc_structures}b).
Furthermore, the simulated 39-particle $B_{26}A_{13}$ cluster is essentially isostructural with the In$_{26}$P$_{13}$ cluster from literature \cite{sandeno_ligand_2024} (Figure~\ref{fig:msc_structures}c), exhibiting the same topology and yielding a bond-angle MAE of 5.50° (see Figs.~S10 and S12 in the Supporting Information). Note that this agreement is achieved without any system-specific parameterization, underscoring that the emergence of complex structural motifs can be captured through generic interparticle interactions alone.

Building on the structural agreement with the experimentally resolved MSCs, we examined both the simulated and experimentally resolved MSCs in detail and found a recurring core motif shared by the non-bulk-like clusters studied here. These clusters are defined by a common $X_{14}Y_{13}$ motif (Figure~\ref{fig:msc_structures}d): one $Y$ atom is located at the center of a tetrahedron of four $X$ atoms, which are encapsulated by a distorted icosahedral cage formed by 12 $Y$ atoms, while the remaining ten $X$ atoms bridge its vertices. By examining the cores and overlapping their atomic positions, we find that they share the same connectivity with only modest geometric distortions (see Figs.~S17 and S18 in the Supporting Information). 
For clarity, we hereafter refer to these non-bulk-like clusters as distorted icosahedral clusters.
On this basis, we next sought to investigate how this geometry gives rise to chirality. 

\begin{figure*}
  \includegraphics[width=\textwidth]{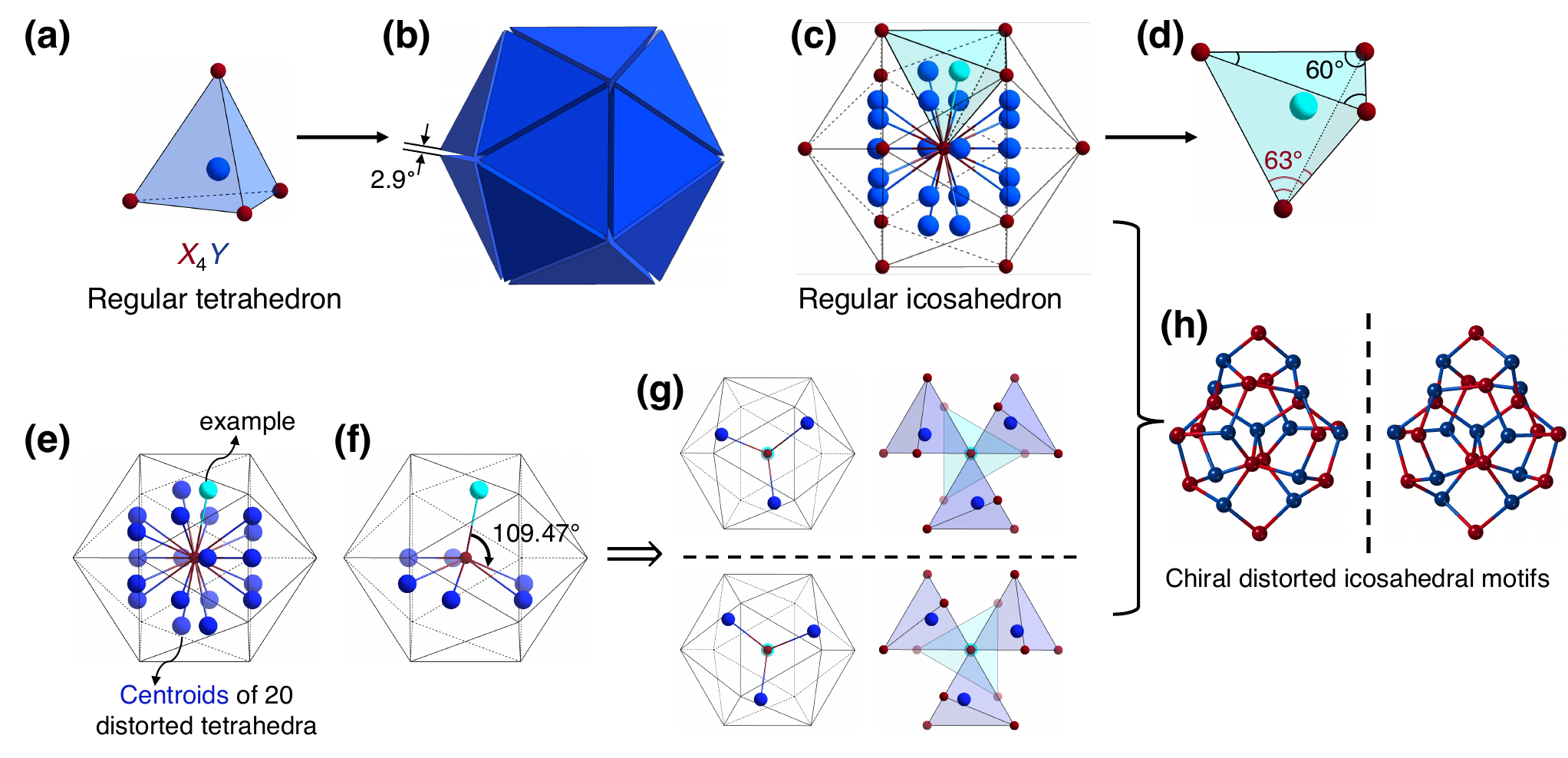}
  \caption{Proposed mechanism for the emergence of chirality in the distorted icosahedral motifs. 
  \textbf{(a)} A regular $X_{4}Y$ tetrahedron can represent the local environment in zincblende- and wurtzite-type structures, but 
  \textbf{(b)} 20 such tetrahedra cannot tile a regular icosahedron without leaving angular gaps of $\sim$2.9°.  
  \textbf{(c--d)} A regular icosahedron can be constructed using 20 slightly distorted tetrahedra, each sharing a common vertex at the icosahedron's center (shown in red). 
  \textbf{(e--f)} Selecting one centroid (cyan) leaves only six of the remaining 19 centroids with which to form ideal tetrahedral bond angles (109.47°). 
  \textbf{(g)} These six centroids can be divided into two sets of three, each forming a tetrahedron with the selected centroid. 
  \textbf{(h)} These two chiral pathways for the cluster construction result in distorted icosahedral motifs with opposite handedness (\textit{i.e.}, related to each other by mirror symmetry).}
  \label{fig:chiral_mechanism}
\end{figure*}

We examined the connection between the icosahedral $X_{14}Y_{13}$ motif and the tetrahedral building blocks in II–VI and III–V semiconductors that adopt zincblende or wurtzite crystal structures in the bulk. While the local environment can be described by regular tetrahedral units $X_{4}Y$ (Figure~\ref{fig:chiral_mechanism}a) that have equivalent $X$--$Y$ bond lengths, twenty regular tetrahedra cannot tile to form a regular icosahedron without leaving angular gaps of about 2.9° between their facets (Figure~\ref{fig:chiral_mechanism}b). Thus, the icosahedral framework is geometrically frustrated, which causes the observed distortions in the $X_{14}Y_{13}$ motif. 
It is, however, possible to construct a regular icosahedron using 20 identical yet slightly distorted tetrahedra (Figure~\ref{fig:chiral_mechanism}c--d), each sharing a common vertex at the icosahedron's center (represented in red in Figure~\ref{fig:chiral_mechanism}e). (Note that the tetrahedra's distortion manifests by the centroids of each distorted tetrahedron---represented in cyan in Figure~\ref{fig:chiral_mechanism}d---not having the same distance to the common vertex in the icosahedron's center as to the other three vertices, which represent the icosahedron's vertices. Concurrently, not all dihedral angles are the same---60°.) 

To inscribe an inner tetrahedron into an icosahedron, one must select one such centroid (represented in cyan in Figure~\ref{fig:chiral_mechanism}e). 
Only six of the remaining 19 centroids form a 109.47° tetrahedral bond angle with the first centroid and the vertex at the center of the icosahedron (Figure~\ref{fig:chiral_mechanism}f).
These six centroids can be divided into two sets of three, each forming a regular tetrahedron with the initially selected centroid (Figure~\ref{fig:chiral_mechanism}g). 

This inscription of a tetrahedral motif into an icosahedral geometry gives rise to two structural motifs with opposite handedness---providing two symmetry-equivalent yet distinct pathways for the assembly of tetrahedral units during cluster formation. 
Because these pathways are statistically equivalent, chiral magic-size clusters are typically synthesized as racemic mixtures in experiments, and they are observed with equal likelihood in simulation.
As the tetrahedron is inscribed into the icosahedron, the symmetry of the icosahedron (point group $\bar{5}{3}m$, Fig.~\ref{fig:chiral_mechanism}e) is reduced first by the addition of the tetrahedral centroids (to point group $m\bar{3}$, Fig.~\ref{fig:chiral_mechanism}f) and further by taking into account the whole tetrahedral network, \textit{i.e.}, the tetrahedron composed of tetrahedra in an icosahedral arrangement (point group $23$, Fig.~\ref{fig:chiral_mechanism}g). 
We propose that the interplay between geometric frustration and symmetry breaking is the origin of chirality in these distorted icosahedral motifs. 
As such, geometric frustration may act as a general principle for the formation of chiral MSCs in II–VI and III–V semiconductor systems.

\begin{figure*}
  \includegraphics[width=\textwidth]{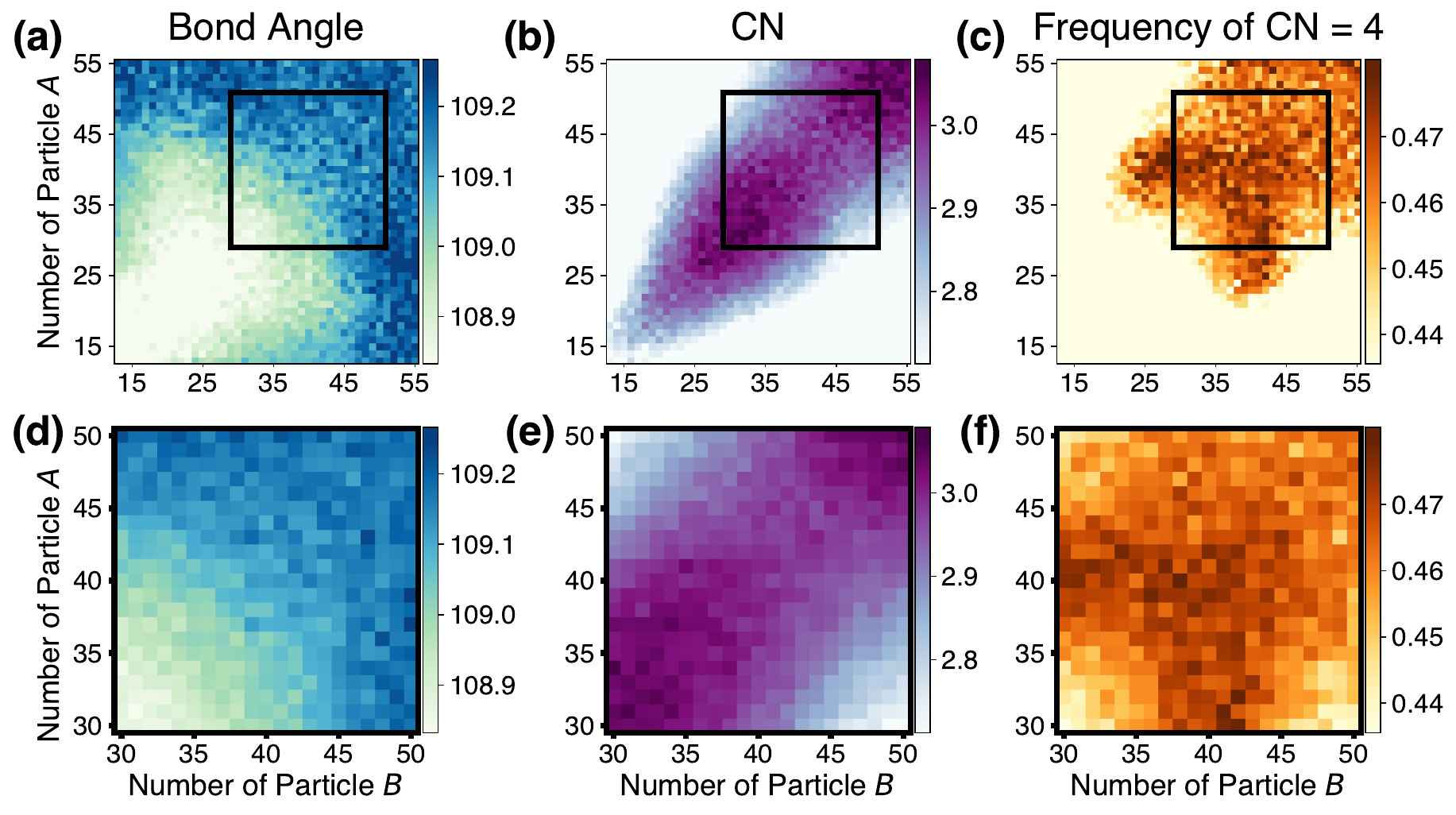}
  \caption{Average bond angles (in degrees) (\textit{left}), average coordination numbers (\textit{middle}), and frequency of optimal coordination CN $= 4$ (\textit{right}) across different stoichiometries and system sizes (stoichiometries are varied in increments of $\Delta N = 1$ between compositions 55:55 and 13:55 / 55:13):
  \textbf{(a--c)} system sizes $N = 26$--110 (10 replica simulations per pixel; black square outlines in each panel highlight the system size range from $N = 60$--100.); 
  \textbf{(d--f)} system sizes $N = 60$--100 (30 replica simulations per pixel; compositions between 50:50 and 30:50 / 50:30). }
  \label{fig:CN_BA}
\end{figure*}

To understand why chiral MSCs preferentially form at specific sizes and compositions, we systematically investigated how system size and stoichiometry influence the self-assembly of nanoclusters. 
In Figure~\ref{fig:CN_BA}, the self-assembled configurations are analyzed using the average bond angle and coordination numbers. 

Clusters across sizes of $N = 26$--110 particles consistently exhibit near-tetrahedral bond angles, approaching 109.47° for increasing cluster sizes (for information on clusters as small as $N = 4$, see Fig.~S21 in the Supporting Information). 
The average bond angles of stoichiometric clusters remain slightly below the optimal tetrahedral angle, but steadily increase with system size (see Figure~\ref{fig:CN_BA}a, diagonally from bottom left to top right, and Fig.~S21c in the Supporting Information); we attribute this to the fact that larger clusters approach bulk-like conditions, where all bond angles can be tetrahedral, whereas a large fraction of particles in smaller clusters are located at the cluster surface, where bond angles are distorted due to dangling ``bonds''. 
Off-stoichiometric clusters, on the other hand, appear to feature optimal bond angles more frequently. 
This behavior likely arises from the excess of one particle type saturating the local bonding environment of the minority species: fewer undercoordinated configurations are observed and local tetrahedral motifs characteristic of the zincblende phase (mimicking \textit{sp}$^3$-hybridization) dominate.

In larger systems ($N \geq 80$), bond angles across the cluster are more uniform and adopt values just below the ideal tetrahedral angle (Figure~\ref{fig:CN_BA}d), consistent with prior X-ray spectroscopic studies of similar semiconductor nanoclusters \cite{liu_analysis_2020}.
Coordination number analysis revealed that average CNs approached the ideal value of 4 near stoichiometric compositions (Figure~\ref{fig:CN_BA}b,e). 
However, the relationship between CN and total particle number is non-monotonic: after an initial rise, average CN values decline between $N \approx 62$ and 88, before increasing again toward $N = 110$ (see Fig.~S19 in the Supporting Information). 
We hypothesize that this is due to a closed-shell-like effect.

Quantifying the number of tetrahedrally coordinated particles (CN = 4) across composition space reveals an arc-shaped region with a particularly large amount of optimal configurations (Figure~\ref{fig:CN_BA}c,f). 
These optimal configurations were localized within a narrow region where the number of $A$-type particles ranged from approximately 38 to 42 and the number of $B$-type particles from 36 to 42, with extensions toward more asymmetric compositions such as ($N_A = 40$, $N_B = 30$) and ($N_A = 30$, $N_B = 41$) (see Fig.~S20 in the Supporting Information). 
Unlike the average CN, which peaks along the stoichiometric axis, the number of ideally coordinated particles exhibits an irregular, off-diagonal trend---likely as a result of the optimality of bond angles being balanced against the optimality of local coordinations, CN.

Overall, for stoichiometric clusters (with $20 \leq N \leq 110$), the average bond angle gradually increases to approach the ideal tetrahedral angle of 109.47° with increasing size. 
The average CN rises rapidly up to $N \approx 60$, decreases slightly toward a minimum around $N \approx 80$--90, and then increases again. As a result, the fraction of ideally coordinated particles increases steadily up to $N \rightarrow 70$ (see Fig.~S22 in the Supporting Information).
Given that optimal bond angles and tetrahedral coordination emerge within different compositional windows, 
the coupling between optimal angular geometry and local coordination environment appears to play a critical role in stabilizing semiconductor MSCs: size-limited MSC assembly seems to occur due to a need to balance optimal bond angles with the optimal number of bonds in the tetrahedral environments of these zincblende-type semiconductor systems. 
The fact that experimentally resolved, icosahedral MSCs occur over a narrow cluster size range reflects a compromise between bond angles and coordination (see Fig.~S23 in the Supporting Information).

\section{Conclusion}

In this study, we show that experimentally reported distorted icosahedral MSCs are intrinsically chiral. By modeling zincblende-forming systems particle-by-particle, we reproduce the bulk structure of zincblende-type semiconductors, follow their transformation as cluster size decreases, and identify the transition to distorted icosahedral geometries. In these distorted icosahedral clusters---characterized by the $X_{14}Y_{13}$ motif---we propose that the structural origin of chirality arises from a combination of geometric frustration and symmetry reduction. 

Remarkably, the structures of our simulated clusters match very well with the experimentally characterized zincblende- and wurtzite-forming semiconductor systems. 
Due to the abstract nature of our computational model, this consistent match with experimental observations corroborates our hypothesis that the specific geometries observed in such semiconductor nano\-clusters originate from a fundamental structure-related geometric mechanism rather than being specific to any particular chemical system. 
By systematically analyzing trends across cluster size and stoichiometry, we further suggest that MSC formation reflects a balance between achieving optimal bond angles and favorable coordination environments. 

By revealing a common structural motif and intrinsic chirality in distorted icosahedral MSCs, our work uncovers fundamental features that have largely gone unrecognized in the field thus far. This insight brings a new level of clarity to the structures of MSCs, advancing the field’s fundamental understanding of how these clusters are formed and stabilized.
Future work should extend these investigations to bulk-like nanoclusters of zincblende-type semiconductor materials.

\begin{acknowledgement}

This material is based upon work supported by the National Science Foundation under Grant No.\ CHE-2003586. 
The authors thank Yuan Yao, Thomas Ugras, River Carson, and Brandi Cossairt for helpful discussions.

\end{acknowledgement}

\begin{suppinfo}
The Supporting Information is available free of charge.
\begin{itemize}
  \item[]Additional information including details on the simulation and model, structural analysis, comparative analysis of experimentally determined cluster geometries reported in the literature, geometrical and topological analyses of the distorted icosahedral motif, structural characteristics depending on system size and stoichiometry, and the thermal stability of chiral cluster structures
\end{itemize}
Simulation results are available at the Materials Data Facility.\cite{Blaiszik2016,dataset}
\end{suppinfo}

\bibliography{achemso}

\end{document}